\documentstyle[12pt]{article}
\newcommand{\be}{\begin{equation}}
\newcommand{\ee}{\end{equation}}
\newcommand{\ba}{\begin{eqnarray}}
\newcommand{\ea}{\end{eqnarray}}
\newcommand{\bi}[1]{\bibitem{#1}}
\newcommand{\fr}[2]{\frac{#1}{#2}}
\newcommand{\non}{\nonumber}
\newcommand{\ar}{\mbox{$\rightarrow$}}

\newcommand{\al}{\mbox{$\alpha$}}
\newcommand{\Z}{\mbox{$Z\alpha$}}
\newcommand{\p}{\mbox{$\vec{p}$}}

\newcommand{\bt}{\mbox{$\beta$}}
\newcommand{\lb}{\left (}
\newcommand{\rb}{\right )}
\newcommand{\la}{\left\langle}
\newcommand{\ra}{\right\rangle}
\newcommand{\z}{\mbox{$\scriptscriptstyle 0 $}}

\begin{document}
\pagestyle{empty}

\hfill BudkerINP -- 94 -- 27

\hfill March 1994

\vspace{1.0 cm}

\begin{center}
{\Large \bf Recoil Correction in the Dirac-Coulomb Problem}
\vspace{2.0 cm}

{\bf A.S. Yelkhovsky}

\vspace{0.5 cm}
Budker Institute of Nuclear Physics,\\
and \\
Physics Department, Novosibirsk University, \\
630090 Novosibirsk, Russia

\vspace{4.0 cm}
\end{center}

\begin{abstract}
The expression for the first recoil correction to the Dirac-Coulomb spectrum
is obtained employing the gauge invariance.
\end{abstract}

\newpage

\pagestyle{plain}
\pagenumbering{arabic}

Relativistic two-body problem in quantum electrodynamics has been exactly
solved in the only limiting case $m/M \ar 0$, \al \ar 0 at fixed \Z\ (here
$M$ and $m$ are masses of the constituents, $Z|e|$ and $e$ are their
electric charges, $\al = e^2/\hbar c$ is the fine structure constant, $\hbar =
c = 1$). In this limit, an infinitely heavy nucleus holds still being the
source of the constant in time Coulomb field. A wavefunction of the system
reduces to that of the light particle, the electron, and obeys the Dirac
equation in the Coulomb field,
\be\label{D}
\lb \vec{\al} \p + \bt m + V_C - E \rb \psi = 0.
\ee

Expressions for the first (linear in $m/M$) recoil corrections to
energies of the Dirac-Coulomb bound states were obtained several years
ago by V.M. Shabaev \cite{1,2}. He used the perturbation theory in \Z ,
summing up contributions of a given order in \Z , linear in $m/M$. The
present note is devoted to a simple derivation of the Shabaev's result,
with only minor reference to the perturbation theory.

As a guiding principle we will use the gauge invariance of QED. To begin
with, let us generalize the equation (\ref{D}) to an arbitrary gauge. Since
$V_C = \Z D_{00}$\footnote{$D_{\mu\nu}$'s make up the photon propagator.},
and an infinitely heavy particle at rest can emit (or absorb) only zero
component of the vector potential, we have:
\be\label{Dag}
\left\{ \al_\mu \lb p_\mu - \Z D_{\mu 0} \rb + \bt m \right\} \psi = 0,
\ee
where $\al_{\z}=1$ by definition, $p_{\z} = E$. The Dirac-Coulomb spectrum,
that is the mutual arrangement of the Green's function singularities at the
complex $E$ plane\footnote{We consider only gauges with $D_{\mu 0}$ constant
in time, so that $E$ is the integral of motion.}, is certainly
gauge-invariant.

Now let us take into account the motion and interaction of the nucleus to
first order in $1/M$. They are described by the term
\be\label{Hh}
\fr{\lb \vec{P} - Z|e| \vec{A} \rb ^2}{2M}
\ee
in the Hamiltonian of the system\footnote{The interaction of the electron
with a nucleus proper magnetic moment, formally of the order $1/M$, is taken
into account straightforwardly, so we do not discuss it here.}. Here
$\vec{P}$ is the operator of a nucleus momentum, while $\vec{A}$ is the
vector potential operator acting at the nucleus site. Due to $M$ in the
denominator, $\vec{A}$ can be taken to act at the origin.

In order to find the first recoil correction to an energy of the electron we
need to average the above expression over the corresponding Dirac-Coulomb
eigenfunction.  There is no problem with the vector potential operator --- it
emits (absorbs) photons which are absorbed (emitted) by the electron.
Difficulties emerge when one tries to determine how the operator
$\vec{P}$ acts on the electron wavefunction. In fact, the simple relation
$\vec{P} = - \p$ holds in the nonrelativistic limit only, when the problem
is truly two-body. The relativistic electron can propagate in both time
directions so that at a fixed time slice one has a number of electrons and
positrons with the total momentum equal to $-\vec{P}$.

To get rid of the problem we will start with the operator
\be\label{A2}
\fr{\lb Ze \vec{A} \rb ^2}{2M},
\ee
whose expectation value can be easily expressed in terms of the known
solution to the Dirac-Coulomb problem. Being only the part of the
nucleus Hamiltonian (\ref{Hh}), the operator (\ref{A2}) is by no means
gauge-invariant. This is also true for its expectation value. The
basic idea is to reconstruct a gauge-invariant expression for the total
energy correction from its known noninvariant part.

Taking the expectation value of (\ref{A2}) over fluctuations of the
electromagnetic field we are left with
\be\label{eq:MM}
\Delta E_{\vec{A}^2} = \fr{(\Z)^2}{M} \int \fr{i\: d q_0}{2\pi}
  \la \al_\mu D_{\mu J} (q_0) G (E-q_0) D_{J\nu} (-q_0) \al_\nu \ra.
\ee
Diagrammatically the right-hand side of this equation is shown in Fig.1.
\begin{center}
 \begin{picture}(100,50)
  \put(18,13){\begin{picture}(64,37)
             \put(64,32){\vector(-1,0){32}}
             \put(32,32){\line(-1,0){32}} \put(28,34){$E-q_0$}
             \multiput(36,0)(8,8){4}{\oval(8,8)[tl]}
             \multiput(36,8)(8,8){4}{\oval(8,8)[br]}
             \multiput(28,0)(-8,8){4}{\oval(8,8)[tr]}
             \multiput(28,8)(-8,8){4}{\oval(8,8)[bl]}
             \put(48,16){\vector(0,-1){0.1}} \put(50,15){$q_0$}
             \put(16,16){\vector(0,1){0.1}} \put(11,15){$q_0$}
             \put(32,0){\circle*{1}}
             \end{picture}}
             \put(23,5){{\it Fig.1.} ``Seagull" contribution to
             the recoil.}
 \end{picture}
\end{center}
The solid line depicts $G$, the Green's function for the Dirac equation in
the Coulomb field, wiggly lines represent photon propagators $D_{\mu J}$ and
$D_{J\nu}$. As far as a gauge is not fixed, it is convenient to make
difference between Lorentz indices corresponding to a nucleus interaction
vertex (upper case) and those corresponding to an electron vertex (lower
case). Overall factor $1/M$ allows us to take the limit $M \ar
\infty$ everywhere else. In particular, the remaining photon propagators
connecting the electron line with the nucleus one and not shown explicitly
in Fig.1, have the upper case index equal to zero (see (\ref{Dag})). As
usual, Latin indices run from 1 to 3, Greek ones run from 0 to 3. For the
sake of brevity and later convenience, only the integral over energy flowing
along the loop is written down explicitly. $E$ in (\ref{eq:MM}) is the
energy of a Dirac-Coulomb eigenstate we average over.

Turn now to the gauge transformation properties of (\ref{eq:MM}).
The above discussion of the Dirac equation in the Coulomb field shows that
$\Delta E_{\vec{A}^2}$ is invariant with respect to a gauge transformation
at the electron site,
\be\label{eT}
\delta D_{\mu \Lambda} = \nabla_{\mu} \varphi_{\Lambda},
\ee
where $\varphi_{\Lambda}$ are arbitrary (linear in time) functions. Hence,
it remains to recover the invariance with respect to the transformation at
the opposite `end' of $D$, attached to the nucleus line, namely
\be\label{nT}
\delta D_{\mu J} = \nabla_{J} \varphi_{\mu}.
\ee
Trying to do this we cannot use components of $D$ with a spatial upper case
index. Actually, those components of $D$ emerge in an expression for the
energy correction due to the operator $\vec{A}$ acting at the nucleus site.
Since the quadratic in $\vec{A}$ effect is already taken into account by
(\ref{eq:MM}), the only thing that can help us reads
\be
\al_{\mu} D_{\mu 0} = D_{\z 0} - \al_l D_{l0}.
\ee
It does not spoil the invariance with respect to (\ref{eT}). On the other
hand, its gauge variation with respect to the evident extension of
(\ref{nT}),
\be
\delta D_{\mu 0}(q_0) = i q_0 \varphi_{\mu},
\ee
does compensate that of $\al_{\mu} D_{\mu J}$ in the linear combination
\be
\al_{\mu} D_{\mu J}(q_0) +
\fr{1}{q_0}\,\al_{\mu} i\left[ \nabla_J, D_{\mu 0}(q_0)\right].
\ee

By this means the gauge-invariant expression for the recoil correction takes
the form:
\ba\label{Etot}
\!\!\!\!\!&&\Delta E = \fr{(\Z)^2}{M} \int \fr{i\: d q_0}{2\pi} \\
\!\!\!\!\!&& \la \al_\mu
\lb D_{\mu J}(q_0) + \fr{1}{q_0} i\left[\nabla_J, D_{\mu 0}(q_0)\right] \rb
G (E-q_0)
\lb D_{J\nu}(-q_0) - \fr{1}{q_0} i\left[ \nabla_J, D_{0\nu}(-q_0) \right] \rb
\al_{\nu} \ra.\non
\ea
Recall that $\al_0 = 1$. Unfortunately, this expression is meaningless until we
define how to treat the new singularity at $q_0 = 0$.

As soon as the gauge invariance is maintained we can choose the mostly
convenient gauge. Without question this is the Coulomb one.  Then the total
energy shift (\ref{Etot}) is naturally broken up into four terms:
\be\label{sum}
\Delta E = \Delta E_{CC} + \Delta E_{CM} + \Delta E_{MC} + \Delta E_{MM},
\ee
the last of which is nothing but the Coulomb gauge version of $\Delta
E_{\vec{A}^2}$ (see (\ref{eq:MM}) and Fig.1), i. e. the
double magnetic exchange contribution to the energy shift. The third and the
second terms comprise the correction arising due to a single magnetic
exchange. Their origin at (\ref{Hh}) is the term $-Z|e|(\vec{P}\vec{A} +
\vec{A}\vec{P}) /2M$. Finally, the pure Coulomb contribution $\Delta E_{CC}$
is just the mean value of the nucleus `kinetic energy' $\vec{P}\,^2/2M$.

To find a prescription according to which the $1/q_0$-singularity should be
passed by one can exploit its independence of \Z\ and analyze the
corresponding expression perturbatively in \Z , to the lowest nontrivial
order. Very natural result for the single-magnetic exchange reads
\be
\fr{1}{q_0} \ar
              \fr{1}{2} \lb \fr{1}{q_0 - i0} + \fr{1}{q_0 + i0} \rb.
\ee
For example, the third term in (\ref{sum}) can be represented
diagrammatically by the sum of two graphs shown in Fig.2. There the
thick line depicts the propagator $(\pm q_0 + i0)^{-1}$ of the infinitely
heavy nucleus having the energy $M \pm q_0$, while the dashed line shows the
interaction through the Coulomb electric field $Z|e| [\vec{\nabla},
D_{00}]$.
\begin{center}
 \begin{picture}(115,50)
  \put(0,13){\begin{picture}(56,37)
             \put(56,32){\vector(-1,0){28}}
             \put(28,32){\line(-1,0){28}} \put(24,34){$E-q_0$}
             \multiput(32,0)(5.33,7.11){5}{\line(3,4){2.6}}
             \multiput(21,0)(-6,8){4}{\oval(6,8)[tr]}
             \multiput(21,8)(-6,8){4}{\oval(6,8)[bl]}
             \put(44,16){\vector(-3,-4){0.1}} \put(47,15){$q_0$}
             \put(12,16){\vector(0,1){0.1}} \put(8,15){$q_0$}
             \thicklines
             \put(32,0){\vector(-1,0){5}}
             \put(28,0){\line(-1,0){4}}
             \end{picture}}
             \put(24,9){$M+q_0$}
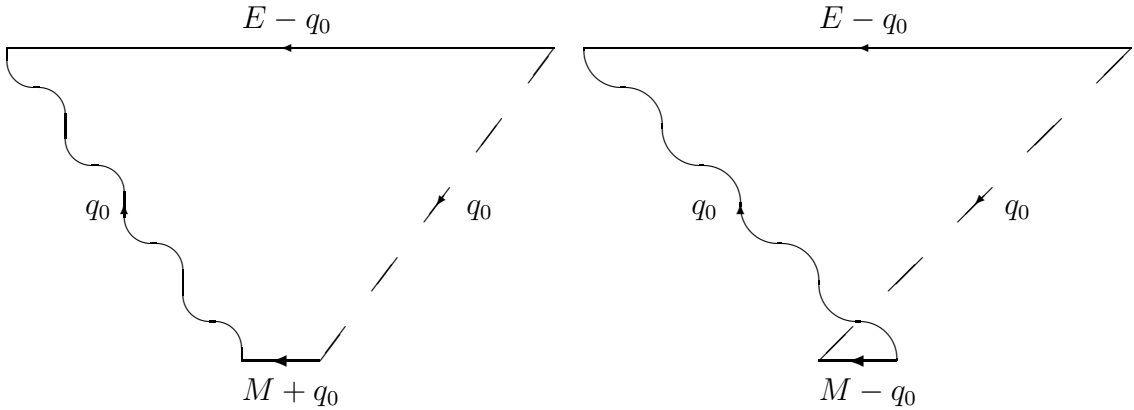
             \put(37,5){{\it Fig.2.} Single magnetic exchange.}
  \put(59,13){\begin{picture}(56,37)
             \put(56,32){\vector(-1,0){28}}
             \put(28,32){\line(-1,0){28}} \put(24,34){$E-q_0$}
             \multiput(24,0)(7.11,7.11){5}{\line(1,1){3.5}}
             \multiput(28,0)(-8,8){4}{\oval(8,8)[tr]}
             \multiput(28,8)(-8,8){4}{\oval(8,8)[bl]}
             \put(40,16){\vector(-1,-1){0.1}} \put(43,15){$q_0$}
             \put(16,16){\vector(0,1){0.1}} \put(11,15){$q_0$}
             \thicklines
             \put(32,0){\vector(-1,0){5}}
             \put(28,0){\line(-1,0){4}}
             \end{picture}}
             \put(83,9){$M-q_0$}
 \end{picture}
\end{center}

Simple perturbative analysis of the pure Coulomb contribution
$\Delta E_{CC}$ suggests that corresponding prescription has the form:
\be\label{p2}
\fr{1}{q_0^2} \ar
              \fr{1}{2} \lb \fr{1}{(q_0 - i0)^2} + \fr{1}{(q_0 + i0)^2} \rb.
\ee
Actually, here we have the sum of the nucleus propagator derivatives
resulting from the expansion of $(\vec{P}\,^2/2M \pm q_0 + i0)^{-1}$.
To obtain Feynman diagrams for the pure Coulomb contribution, one need only
substitute wiggly lines in Fig.2 for dashed ones as well as the nucleus
propagator for its square.

Now, when the integral in (\ref{Etot}) is completely defined one can easily
check that this expression equals the sum of recoil corrections found by
Shabaev in the more straightforward way \cite{1,2} (in \cite{2}, the overall
sign of the expression for $\Delta E_{MM}$ is corrected).

Recall that our starting point was the expectation value (\ref{eq:MM}). Its
perturbative expansion can be readily appreciated to begin with $(\Z)^5$. To
be certain that the corresponding expansion for the total correction
(\ref{Etot}) begins with $(\Z)^2$, let us consider in greater detail the pure
Coulomb contribution which alone survives the transition to the
nonrelativistic limit. As a byproduct we will see how the expectation
value $\la \vec{P}\,^2/2M \ra$ looks in terms of the solution to the
Dirac-Coulomb problem.

Evaluating the integral with respect to $q_0$ in $\Delta E_{CC}$ according
to the prescription (\ref{p2}) together with the standard rules for the
Dirac-Coulomb Green's function one readily obtains
\be\label{CC}
\Delta E_{CC} = \fr{1}{2M} \la \p \lb \Lambda_+ - \Lambda_- \rb \p \ra,
\ee
where $\Lambda_{\pm}$ are the projection operators to sets of positive- and
negative-energy Dirac-Coulomb eigenstates correspondingly. Passing from
(\ref{Etot}) to (\ref{CC}) we used the Dirac equation (\ref{D}). In the
nonrelativistic limit $\Lambda_+ \ar 1$, $\Lambda_-\ar 0$
and (\ref{CC}) reduces to the well-known result,
\be
\Delta E \ar \la \fr{\p\,^2}{2M} \ra.
\ee

\bigskip

{\bf Acknowledgements}

I thank I.B. Khriplovich, A.I. Milstein and M.E. Pospelov for stimulating
discussions. The partial support from the program ``Universities of Russia''
is gratefully acknowledged.

\end{document}